\documentclass[12pt,preprint]{aastex}

\usepackage{aas_macros}
\usepackage{natbib}
\usepackage{epsfig,psfrag,graphics,verbatim}
\usepackage{graphicx}
\usepackage{dcolumn}
\usepackage{bm}
\usepackage{color}

\begin{document}


\title{The dark matter profile of the Milky Way:\\ a non-parametric reconstruction}


\author{Miguel Pato}
\affil{The Oskar Klein Centre for Cosmoparticle Physics,\\ Department of Physics, Stockholm University, AlbaNova, SE-106 91 Stockholm, Sweden}
\affil{Physik-Department T30d, Technische Universit\"at M\"unchen, James-Franck-Stra\ss{}e, D-85748 Garching, Germany}
\and
\author{Fabio Iocco}
\affil{ICTP South American Institute for Fundamental Research, and Instituto de F\'isica Te\'orica - Universidade Estadual Paulista (UNESP), Rua Dr.~Bento Teobaldo Ferraz 271, 01140-070 S\~{a}o Paulo, SP Brazil}
\affil{Instituto de F\'isica Te\'orica UAM/CSIC,\\ C/ Nicol\'as Cabrera 13-15, 28049 Cantoblanco, Madrid, Spain}



\begin{abstract}
\vspace{-14cm}
\begin{flushright}
\small TUM-HEP 976/15
\end{flushright}
\vspace{13.cm}
We present the results of a new, non-parametric method to reconstruct the Galactic dark matter profile directly from observations. Using the latest kinematic data to track the total gravitational potential and the observed distribution of stars and gas to set the baryonic component, we infer the dark matter contribution to the circular velocity across the Galaxy. The radial derivative of this dynamical contribution is then estimated to extract the dark matter profile. The innovative feature of our approach is that it makes no assumption on the functional form nor shape of the profile, thus allowing for a clean determination with no theoretical bias. We illustrate the power of the method by constraining the spherical dark matter profile between $2.5$ and $25\,$kpc away from the Galactic centre. The results show that the proposed method, free of widely used assumptions, can already be applied to pinpoint the dark matter distribution in the Milky Way with competitive accuracy, and paves the way for future developments.
\end{abstract}

\section{Introduction}
\par Mapping out the distribution of dark matter in our Galaxy rests as a paramount task with potentially far reaching implications for astroparticle physics and cosmology. This  is important to understand galaxy formation and to feed searches aimed at unveiling the very nature of dark matter. In particular, direct and indirect dark matter searches rely heavily on the findings of numerical simulations. It is therefore essential to extract the Galactic dark matter distribution directly from observations. In the outer Milky Way (at Galactocentric radii greater than $\sim 20\,$kpc), where baryons contribute little to the total mass budget, the gravitational potential traces closely the dark matter component and the total mass enclosed can be constrained using convenient tracers \citep[e.g.,][]{Sakamoto:2002zr,2006MNRAS.369.1688D,Xue:2008se,Bhattacharjee2014,2014ApJ...794...59K}, although with important degeneracies in the tracer population modelling. By contrast, in the inner Galaxy (i.e., in the inner $\sim 20\,$kpc) the baryonic contribution is very significant and its morphology rather uncertain, which makes the evidence for dark matter difficult to establish and the extraction of its distribution a delicate undertaking \citep{2015NatPh..11..245I}. This has been addressed by many authors with different methods \citep[e.g.,][]{Dehnen:1996fa,Sofue2009,CatenaUllio2010,Weber:2009pt,Iocco:2011jz,Nesti:2013uwa,Bovy:2013raa,2014ApJ...794..151L}, all of which do, however, make explicit assumptions about the underlying dark matter profile: typically, a multi-parameter profile is fitted to the observations together with a given baryonic component. The class of ``local'' methods to measure the dark matter density in the solar neighbourhood \citep[e.g.,][]{Salucci:2010qr,Smith:2011fs,Garbari:2012ff,Zhang:2012rsb,Read:2014qva} avoids this bias, yet such methods are not easily applicable elsewhere in the Galaxy. An approach free of profile assumptions has been developed and successfully tested in external galaxies \citep{Persic:1995ru,Salucci:2007tm}, but never applied to our own Galaxy given the sizeable uncertainties of both kinematic data and baryonic modelling.

\par In this Letter we show that the latest rotation curve measurements and baryonic models make it possible to infer the dark matter profile directly from Milky Way observations without unnecessary assumptions. Our results confirm that this approach is quantitatively competitive to the others used so far, while presenting the noticeable advantage of making no a priori assumption on how dark matter is distributed across the Galaxy.

\section{Methodology}
\par The total gravitational potential of our Galaxy can be written as a sum of two components, namely baryons and dark matter: $\phi_{tot}=\phi_{b}+\phi_{dm}$. The left-hand side (or rather its radial derivative) is fixed by measurements of the rotation curve, whereas the first term on the right-hand side is set by the observed distribution of stars and gas. These are the two data inputs needed to infer the distribution of Galactic dark matter.

\par Regarding the rotation curve, we determine the angular circular velocity $\omega_c$ with a broad collection of tracers comprising gas kinematics (HI and CO terminal velocities, HI thickness, HII regions, giant molecular clouds; \citet{Fich1989,McClure-GriffithsDickey2007,Luna2006,HonmaSofue1997,BrandBlitz1993,Hou2009}), star kinematics (open clusters, planetary nebulae, classical cepheids, carbon stars; \citet{FrinchaboyMajewski2008,Durand1998,Pont1997,Battinelli2013}) and masers \citep{Reid2014} in a total of 2780 measurements distributed across Galactocentric radii $R=0.5-25\,$kpc. Our compilation of data improves upon commonly used compilations \citep{Sofue2009,Bhattacharjee2014} by including numerous tracers available in the literature but often neglected. For each object, the measured line-of-sight velocity in the local standard of rest $v_{lsr}^{los}$ is converted into the angular circular velocity through $v_{lsr}^{los} = \left( R_0\omega_c - v_0  \right) \cos b \, \sin \ell$, where $\ell,b$ are the Galactic longitude and latitude, $R_0$ is the distance to the Galactic centre and $v_0\equiv v_c(R_0)$ the local circular velocity. Throughout the analysis we take $R_0=8\,$kpc, $v_0=230\,$km/s and the peculiar solar motion of \citet{Schoenrich2010}. The uncertainties on both distance and kinematics are assigned according to each source reference and propagated to $R$ and $\omega_c$, respectively. We have checked that our determination of the rotation curve is solid against systematics due to spiral arms \citep{BrandBlitz1993} and against the non-circularity of tracer orbits. Note that $\omega_c$ is used instead of the actual circular velocity $v_c\equiv R\,\omega_c$ since the errors of $\omega_c$ and $R$ are not correlated, while those of $v_c$ and $R$ are. The total acceleration is then given by $\textrm{d}\phi_{tot}/\textrm{d}R = \omega_c^2 R$.

\par For the baryonic component, we implement a wide range of alternative observation-based distributions for the stellar bulge \citep{Stanek1997,Zhao1996,BissantzGerhard2002,LopezCorredoira2007,Vanhollebeke2009,Robin2012}, stellar disc(s) \citep{HanGould2003,CalchiNovatiMancini2011,deJong2010,Juric2008,Bovy:2013raa} and gas \citep{Ferriere1998,Moskalenko2002}. The bulge models comprise different parameterizations for the Galactic bar and are normalised to microlensing optical depth data \citep{MACHO2005} using a procedure thoroughly described in \citet{Iocco:2011jz}. The discs bracket a variety of profiles fitted to photometric observations and are calibrated to the latest measurement of the local total stellar surface density \citep{Bovy:2013raa}. The two alternative gas models adopted \citep{Ferriere1998,Moskalenko2002} consist of molecular, atomic and ionised phases whose spatial distributions have been traced by wide surveys of different spectral lines (mainly $21\,$cm and CO); we have checked that using the more massive HI disc of \citet{2008A&A...487..951K} has no significant impact on our results. The contribution of each component to the rotation curve is computed through multipole expansion \citep{BinneyTremaine} and the statistical uncertainty on its normalisation is propagated accordingly. By summing in quadrature the contribution of the three components (bulge, disc, gas) in all possible combinations, we obtain a compilation that covers the entire range of morphologies available in the literature, thus bracketing the baryonic contribution to the total acceleration, $\textrm{d}\phi_{b}/\textrm{d}R=\omega_{b}^2 R$. It is crucial to notice here that, to the best of our knowledge, our library of observation-inferred baryonic distributions includes virtually all morphologies available in the literature, which allows us to adopt the resulting spread as an estimate of the systematic uncertainty on baryonic modelling rather than use a single model with unknown systematics.

\par
\begin{figure*}[ht]
\begin{center}
\includegraphics[width=1.\textwidth]{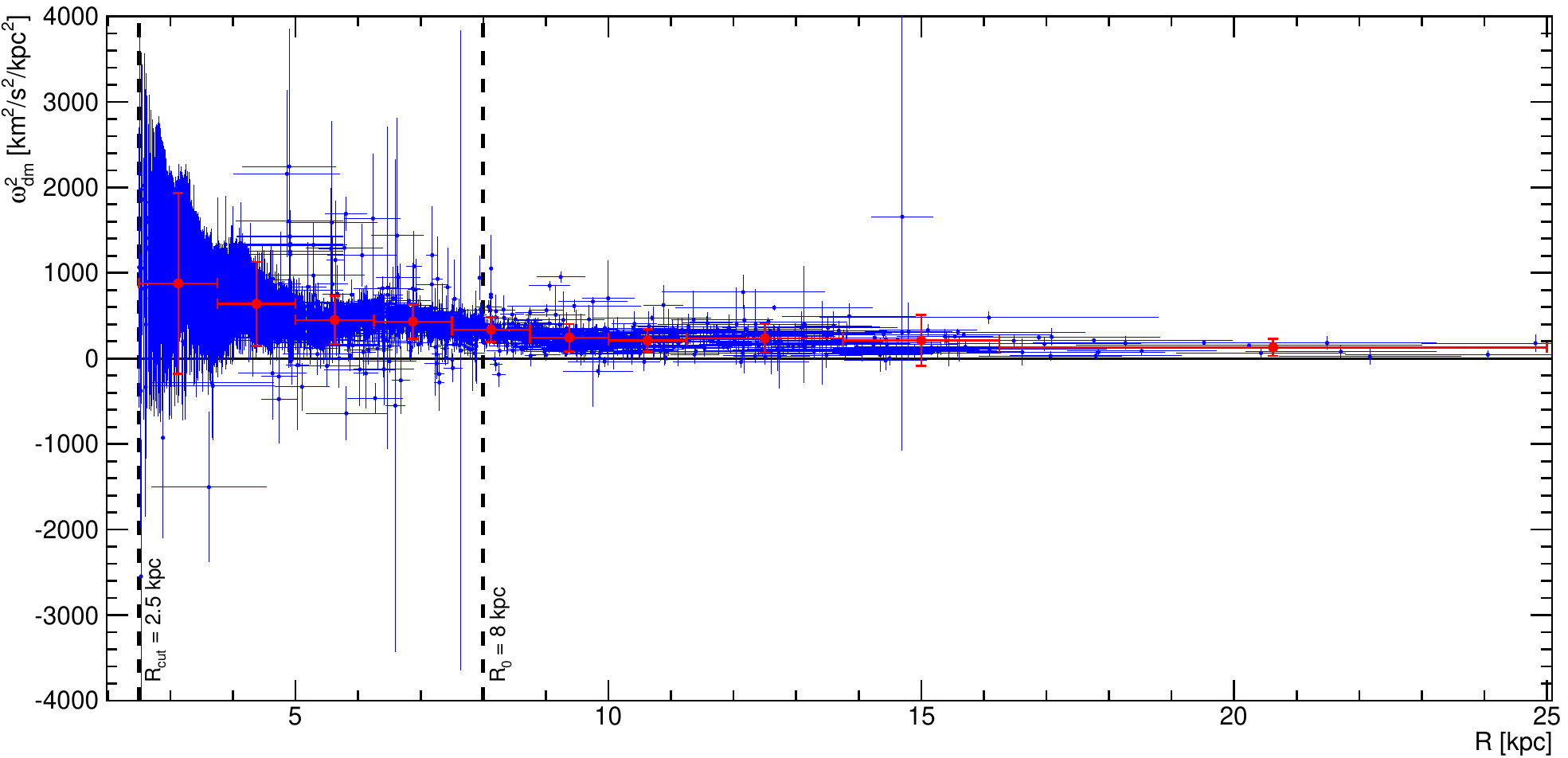}
\caption{The dark matter contribution to the rotation curve of our Galaxy. The blue data points show the dark matter residuals $\omega_{dm}^2$ as inferred from the latest compilation of rotation curve measurements and for the representative baryonic model \citep{Stanek1997,Bovy:2013raa,Ferriere1998}. The red data points display the residuals after applying the binning procedure described in the text. The horizontal bars in the binned residuals are solely to indicate the length of the radial bins. Note that the dark matter contribution to the actual rotation curve $v_c$ reads $v_{dm}=R\omega_{dm}$.}
\label{fig:residuals}
\end{center}
\end{figure*}

 The core aim of this work is to subtract $\omega_{b}$ from $\omega_c$. Defining $\textrm{d}\phi_{dm}/\textrm{d}R = \omega_{dm}^2 R$, the decomposition of the potential implies
\begin{equation}\label{eq:resid}
\omega_{dm}^2=\omega_c^2-\omega_{b}^2
\end{equation}
under the assumption that the discrepancy between the observed rotation curve and that expected from the distribution of observed baryons is caused by an underlying dark matter component. The inferred residuals $\omega_{dm}^2$ and the corresponding uncertainties (propagated from both $\omega_c$ and $\omega_{b}$) are shown in Fig.~\ref{fig:residuals} for a baryonic model comprising a specific bulge \citep{Stanek1997}, disc \citep{Bovy:2013raa} and gas \citep{Ferriere1998}, chosen for representative purposes as it lies close to the median value of the baryonic envelope at all $R$ (see next section). All objects within $2.5\,$kpc from the Galactic centre are omitted to avoid tracers with non-circular orbits. The residuals are consistently above zero and grow towards the centre. These data trace $\textrm{d}\phi_{dm}/\textrm{d}R$ and not directly the dark matter density distribution $\rho_{dm}$. However, the radial slope of $\omega_{dm}^2\equiv v_{dm}^2/R^2$ does contain information about $\rho_{dm}$. For simplicity, let us take a spherically symmetric dark matter component. Then, one obtains the well-known relation $v_{dm}^2 = G M_{dm}(<\!\!R)/R$, which can be easily solved for the density in a spherical shell of radius $R$:
\begin{eqnarray}\nonumber
\rho_{dm} &=& \frac{1}{4\pi G}\left( 3\,\omega_{dm}^2 + R\,\frac{\textrm{d}\omega_{dm}^2}{\textrm{d}R} \right) \\ \label{eq:master}
                   &=& \frac{\omega_{dm}^2}{4\pi G}\left( 3 + \frac{\textrm{d}\ln\omega_{dm}^2}{\textrm{d}\ln R} \right) \quad ,
\end{eqnarray}
where the last step is strictly valid only for $\omega_{dm}^2\!>\!0$. The same result follows directly from the Poisson equation for a spherical potential. Effectively, any deviation from the scaling $\omega_{dm}^2 \propto R^{-3}$ indicates the presence of dark matter and the magnitude of such deviation is a measure of its density at radius $R$. For the emblematic case of a flat rotation curve $v_{dm} \simeq v_c = \textrm{const}$, the usual scaling $\rho_{dm}\propto R^{-2}$ is recovered. Eq.~(\ref{eq:master}) is our master formula to extract the (spherical) dark matter profile directly from the data. Notice that (i) no assumption has been made about the functional form of $\rho_{dm}(R)$, and (ii) in principle it is possible to find the equivalent of Eq.~(\ref{eq:master}) for non-spherical geometries by solving $\textrm{d}\phi_{dm}(\rho_{dm})/\textrm{d}R=\omega_{dm}^2 R$.

\par The determination of the profile requires an estimate of $\omega_{dm}^2$ and its radial slope. The first step adopted is to bin the data, which comprises 2687 individual measurements in the range $R=2.5-25\,$kpc. We start by setting up 18 linearly spaced intervals in this range and then merge adjacent intervals as necessary to have at least five measurements per bin and a mean uncertainty in $R$ smaller than the bin half-width. Next, we compute the simple average of $\omega_{dm}^2$ in each bin and take for its uncertainty the quadrature of the mean of $\omega_{dm}^2$ uncertainties and the standard deviation of the central values. Finally, data points more than five sigma away from the $\omega_{dm}^2$ average are excluded and the procedure is repeated until convergence. Typically, about 12 outliers are excluded from the initial set, and the remaining measurements are distributed into 10 radial bins. The number of measurements per bin range from 19 (for the outermost bin) to around 575. The resulting binned residuals are shown in red in Fig.~\ref{fig:residuals} and are used to set $\omega_{dm}^2$ in Eq.~(\ref{eq:master}). 

\par The next step is to determine the slope of $\omega_{dm}^2$, namely the second term in Eq.~(\ref{eq:master}). Instead of using  the average residuals explained above (which would lead to unnecessary correlations and overestimated uncertainties), we use the individual data points within each bin. Specifically, in what we call our default method and for which the final results are shown, a weighted straight-line fit of the points in each bin is performed to estimate $\textrm{d}\omega_{dm}^2/\textrm{d}R$ and the corresponding uncertainty. In order to validate the radial slope estimates of our default method, we have implemented three alternative procedures. In the first, a proxy of the default method, the measurements within each bin were fitted with a power law (rather than a straight line) to determine the logarithmic slope in the second line of Eq.~(\ref{eq:master}). Separately, the slope $\textrm{d}\omega_{dm}^2/\textrm{d}R$ has been estimated as a differential between adjacent bins with two different methods. The first relies simply on the difference between the average values of $\omega_{dm}^2$ across adjacent bins. The second is slightly more involved: $\textrm{d}\omega_{dm}^2/\textrm{d}R$ has been computed for each single data point in every bin as an average of the differentials obtained with all the data points of the previous bin. The slope $\textrm{d}\omega_{dm}^2/\textrm{d}R$ assigned to the bin is then the average of all single-point values, and uncertainties are computed as a spread around that central value. The last two methods (based on the adjacent bin differential) present the disadvantage of having highly correlated uncertainties. However, the central values for the slope obtained with all four methods are in remarkable agreement, as discussed in the next section.

\begin{figure*}[h]
\begin{center}
\includegraphics[width=1.\textwidth]{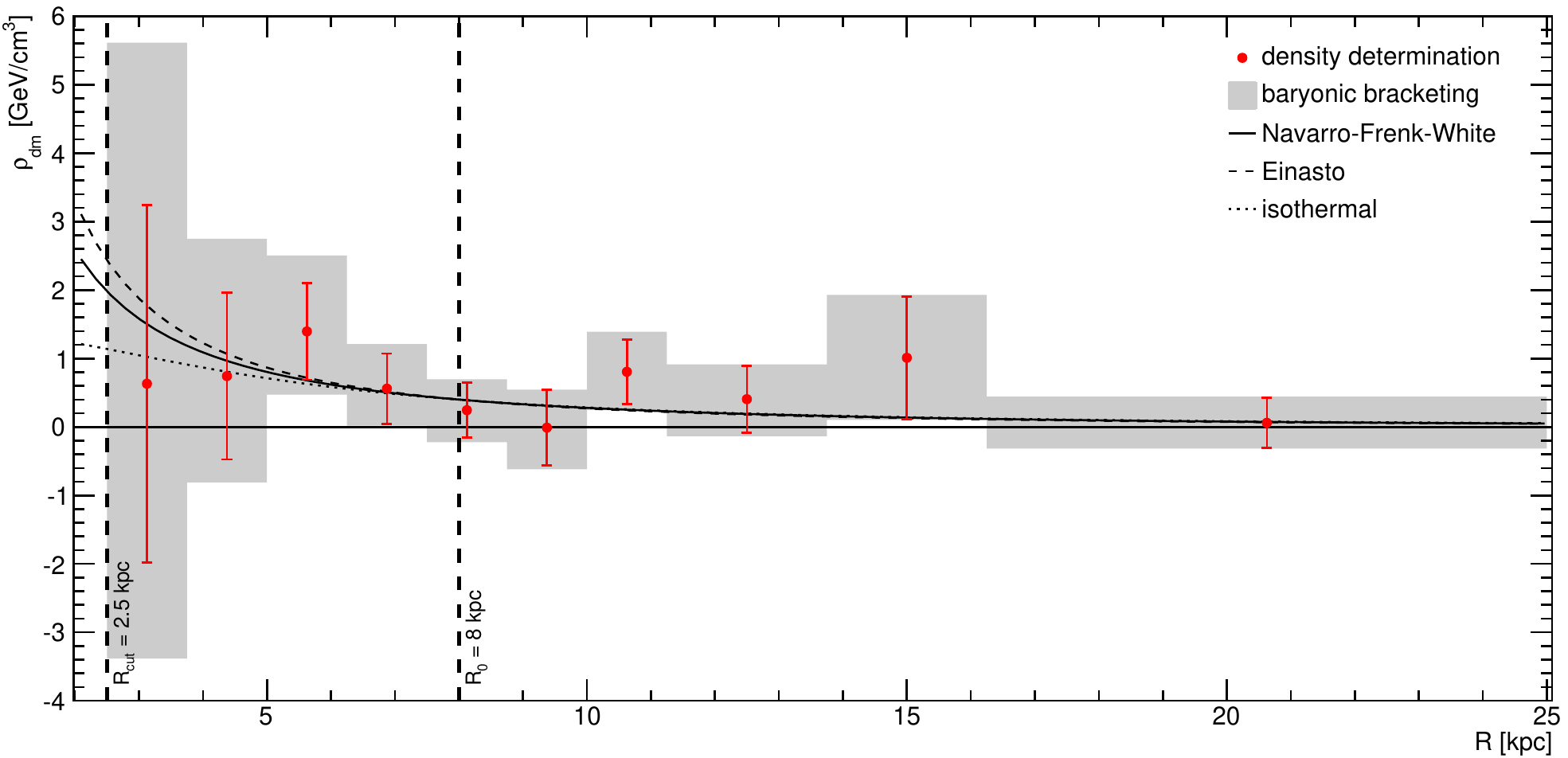}
\caption{The spherical dark matter profile of our Galaxy as inferred directly from observations. The red data points represent the one sigma measurement of the profile for the radial binning shown in Fig.~\ref{fig:residuals} and the representative baryonic model \citep{Stanek1997,Bovy:2013raa,Ferriere1998}. The grey region encompasses the results for all baryonic models, including the corresponding one sigma uncertainties. Overplotted in black are commonly used dark matter profiles, namely Navarro-Frenk-White with scale radius $r_s=20\,$kpc (solid line), Einasto with $r_s=20\,$kpc and slope parameter $\alpha=0.17$ (dashed) and cored isothermal with $r_s=5\,$kpc (dotted), all normalised to a local dark matter density of $0.4\,$GeV/cm$^3$. For reference, $0.38\,$GeV/cm$^3 = 0.01\,$M$_\odot$/pc$^3$.}
\label{fig:prof}
\end{center}
\end{figure*}

\section{Results}
\par We now have all the necessary ingredients to determine the dark matter profile using Eq.~(\ref{eq:master}). This was done across Galactocentric radii $R=2.5-25\,$kpc for each baryonic model. We thus obtain an envelope for $\rho_{dm}$ which encompasses both the statistical uncertainty arising from the residual $\omega_{dm}^2$ and its slope, and the systematic uncertainty due to our ignorance of the actual morphology of stars and gas in the Galaxy. Fig.~\ref{fig:prof} shows the determination of the profile obtained by applying the default method to compute the radial slope $\textrm{d}\omega_{dm}^2/\textrm{d}R$. The error bars represent the one sigma uncertainties for the representative baryonic model; note that these uncertainties result from the propagation of errors on the rotation curve $\omega_c$, the normalisation of the baryonic component $\omega_b$, the Galactocentric radius $R$ and the slope $\textrm{d}\omega_{dm}^2/\textrm{d}R$. The grey region encompasses the one sigma determination for all baryonic models implemented.

\par Before discussing Fig.~\ref{fig:prof}, a few comments are in order regarding the robustness of our findings. Firstly, we note that the four methods devised to estimate the radial slope (see previous section) are all compatible with each other at the one sigma level, thus speaking for the consistency of the presented profile determination. Moreover, we have explicitly checked that the results are solid against the choice of the radial binning and the non-exclusion of outlier data points. Notice that throughout this work $R_0=8\,$kpc and $v_0=230\,$km/s; adopting different values does not qualitatively change our conclusions.

\par The results in Fig.~\ref{fig:prof} present several remarkable features. In order to address these properly, it is important to point out that the first piece in Eq.~(\ref{eq:master}) dominates over the second one across the whole range of Galactocentric radii addressed here. The slope of the dark matter residuals is therefore sub-leading (but not negligible) in the determination of $\rho_{dm}$. We first comment on the magnitude of the uncertainties shown in Fig.~\ref{fig:prof}. In the innermost bins, both the uncertainty for a single baryonic model and the dispersion due to baryonic modelling are large. This is because baryons dominate the gravitational potential below about $5\,$kpc. On the one hand, the little room left for a dark matter contribution to the rotation curve prompts the considerable uncertainty of the reconstructed $\rho_{dm}$ for each baryonic model. On the other hand, the leading role of baryons in this region gives weight to a broad dispersion in the contribution of different morphologies of the inner Galaxy, which in turn shows up in the extended size of the grey region below about $5\,$kpc. Such effect is mitigated between $6$ and $10\,$kpc, where the baryons give a decreasingly important contribution to the gravitational potential. In that intermediate region, the mild uncertainties reported are dominated by the dispersion of the rotation curve data. This dispersion then grows towards larger radii, causing the fluctuations seen above $10\,$kpc.

\par The uncertainties in the innermost regions will eventually be improved with data soon to be provided by Gaia \citep{deBruijne:2012xj}, whose dramatic impact on the census of the Galaxy will very likely help reduce the spread on the current models of the inner Milky Way. For larger Galactocentric radii, an increase in number and precision of rotation curve measurements (or the use of alternative kinematic tracers) would also improve the reconstruction of the dark matter profile. Interestingly enough, although our method is not optimised to measure the dark matter density in the solar neighbourhood, we do find a density at $R\simeq R_0$ close to the usual values $0.3-0.5\,$GeV/cm$^3$ obtained by both global \citep{Dehnen:1996fa,Sofue2009,CatenaUllio2010,Weber:2009pt,Iocco:2011jz,Nesti:2013uwa,Bovy:2013raa} and local \citep{Salucci:2010qr,Smith:2011fs,Garbari:2012ff,Zhang:2012rsb,Read:2014qva} methods. Notice in particular that our results are compatible with the findings of \citet{Salucci:2010qr} even though we use no constraints on the Oort's constants to fix the local slope of the rotation curve. Furthermore, as shown in Fig.~\ref{fig:prof}, our reconstructed profile is in agreement with those inferred from numerical simulations \citep{NFW1996,Merritt2006}, but current uncertainties hinder any discrimination power between different radial behaviours. In principle, it would be possible to shrink the reported uncertainties at the expense of forcing a generic functional form for the dark matter profile (e.g., a monotonicity prior), but we refrain to do so here in order not to spoil the innovative feature of our technique. Our minimal approach also allows for future observational tests of theoretical priors, thus making it a powerful diagnostic tool.

\section{Conclusion}
\par It is truly remarkable that, despite decades of observations and theoretical progress, the distribution of dark matter in the Milky Way remains largely unconstrained. The situation is particularly problematic towards the inner Galaxy, where baryons dominate the gravitational potential and for which any solid improvement would have a remarkable impact on astroparticle physics and cosmology. In this context, we have for the first time implemented a method to reconstruct the Galactic dark matter profile directly from observations. The method requires no assumption on the form or shape of the profile, unlike all previous techniques applied to the Milky Way. Our findings -- obtained using the most recent kinematic data and baryonic models -- are in good agreement with numerical simulations and with both local and global measurements of the local dark matter density. These results can be improved both on the observational side (e.g., by including future Gaia data) and on the theoretical side (e.g., by relaxing the assumption of spherical symmetry). The proposed technique, complementary and competitive to others in the literature, represents a step forward towards achieving a more accurate description of the dark matter distribution in our Galaxy.


\acknowledgments
We thank G.~Bertone and P.~D.~Serpico for useful comments on the manuscript. M.~P.~acknowledges the support from Wenner-Gren Stiftelserna in Stockholm. F.~I.~acknowledges the support of the Spanish MINECO’s ``Centro de Excelencia Severo Ochoa'' Programme under grant SEV-2012-0249 and the Consolider-Ingenio 2010 Programme under grant MultiDarkCSD2009-00064. Part of this work has been carried out during the workshop ``{\it What is the Dark Matter?}'' at NORDITA, Stockholm.


\bibliographystyle{apj}                       
\bibliography{profreconst}


\end{document}